\documentclass{statsoc}
\usepackage[a4paper]{geometry}
\usepackage{lineno,hyperref}
\usepackage{changebar}
\usepackage{soul}
\usepackage{verbatim}
\usepackage{natbib}
\newcounter{algno}
\def\mywidth{0.7\textwidth}

\newenvironment{algorithm}[1]{
\bigskip
\noindent
\rule{\textwidth}{1mm}
\begin{footnotesize}
{\tt #1:}
}
{
\end{footnotesize}
\noindent
\rule{\textwidth}{1mm}
}

\modulolinenumbers[5]

\pdfoutput=1
\usepackage{amsmath,amsfonts,amssymb}
\usepackage{graphicx}
\usepackage{subfigure}
\def\beq{\begin{equation}}
\def\eeq{\end{equation}}
\newcommand{\bm}[1]{\mbox{\protect\boldmath $ #1 $}}
\def\E{{\mathbb E}}
\newcommand{\cov}[1]{\, {\rm Cov}\left( #1 \right) }
\newtheorem{definition}{Definition}[section] %for article style
\newcommand{\ess}{\operatorname{ess}}
\newcommand{\p}[2]{\, q\left( #1, #2 \right) }
\def\ph{\varphi}
\newcommand{\phis}{\ph_{[zT],\tau,\tau}}
\newcommand{\bb}{\bm{b}_{[zT]}^{(b)}}

\renewcommand{\bf}{\bm{b}_{[zT]+\tau}^{(f)}}

\newcommand{\BT}{{\rm B}_{[zT]}}
\newcommand{\Bb}{{\rm B}_{[zT]}^{(b)} }

\newcommand{\Bf}{{\rm B}_{[zT]+\tau}^{(f)}}

\newcommand{\PPhis}{\bm{\ph}_{[zT],\tau}}
\newcommand{\rhs}{\, \bm{r} }

\newcommand{\sigb}{\Sigma_{[zT];T}^{(b)}}
\newcommand{\sigf}{\Sigma_{[zT]+\tau;T}^{(f)}}

\newcommand{\mmspe}{\operatorname{MSPE}}

\newcommand{\fryz}{\operatorname{FVBvS}}
\newcommand{\us}{\operatorname{FORLAP}}

\def\bcb{\begin{changebar}}
\def\ecb{\end{changebar}}
\usepackage{color,xcolor}

\setstcolor{red}

\graphicspath{{./Plots/}}

\title[Forecasting Gross Value Added]{Automatic Locally Stationary Time Series Forecasting with application to predicting U.K. Gross Value Added Time Series under sudden shocks caused by the COVID pandemic}

\author[R. Killick {\it et al.}]{Rebecca Killick}
\address{Lancaster University,
Lancaster,
United Kingdom.}

\author{Marina I. Knight}
\address{University of York,
York,
United Kingdom.}

\author{Guy P. Nason}
\address{Imperial College London,
London,
United Kingdom.}

\author{Matthew A. Nunes}
\coaddress{Matthew A. Nunes, Department of Mathematical Sciences, University of Bath,
Bath, BA2 7AY, UK.\\ Email: m.a.nunes@bath.ac.uk
}
\address{University of Bath,
Bath,
United Kingdom.}

\author[R. Killick {\it et al.}]{Idris A. Eckley}
\address{Lancaster University,
Lancaster,
United Kingdom.}

\begin{document}

\begin{abstract}
Accurate forecasting of the U.K. gross value added (GVA) is fundamental for measuring the growth of the U.K. economy.  A common nonstationarity in GVA data, such as the ABML series,
is its increase in variance over time due to inflation. Transformed or inflation-adjusted series can still be challenging for classical stationarity-assuming forecasters. We adopt a different approach that
works directly with the GVA series by advancing recent forecasting methods for locally stationary time series. Our approach results in more accurate and reliable forecasts, and continues to work well
even when the ABML series becomes highly variable during the COVID pandemic.
\end{abstract}


\section{Introduction}\label{sec:data}

The literature on forecasting stationary time series has been established for many years. See, for example, \cite{gardner85:exponential} or \cite{box70:time} with easily implemented code readily available on a variety of platforms. Rather surprisingly, the same cannot be so readily said when it comes to forecasting of nonstationary time series. Indeed, it is not uncommon for analysts to forecast time series assuming, but not testing for, second-order stationarity. Yet, as \cite{janeway09:six} describes, there can be grave consequences for ignoring this nonstationary structure. This article seeks to address this by providing practical, implemented forecasting methods for nonstationary series.

\begin{figure}[!h]
\centering
\resizebox{\mywidth}{!}{\includegraphics{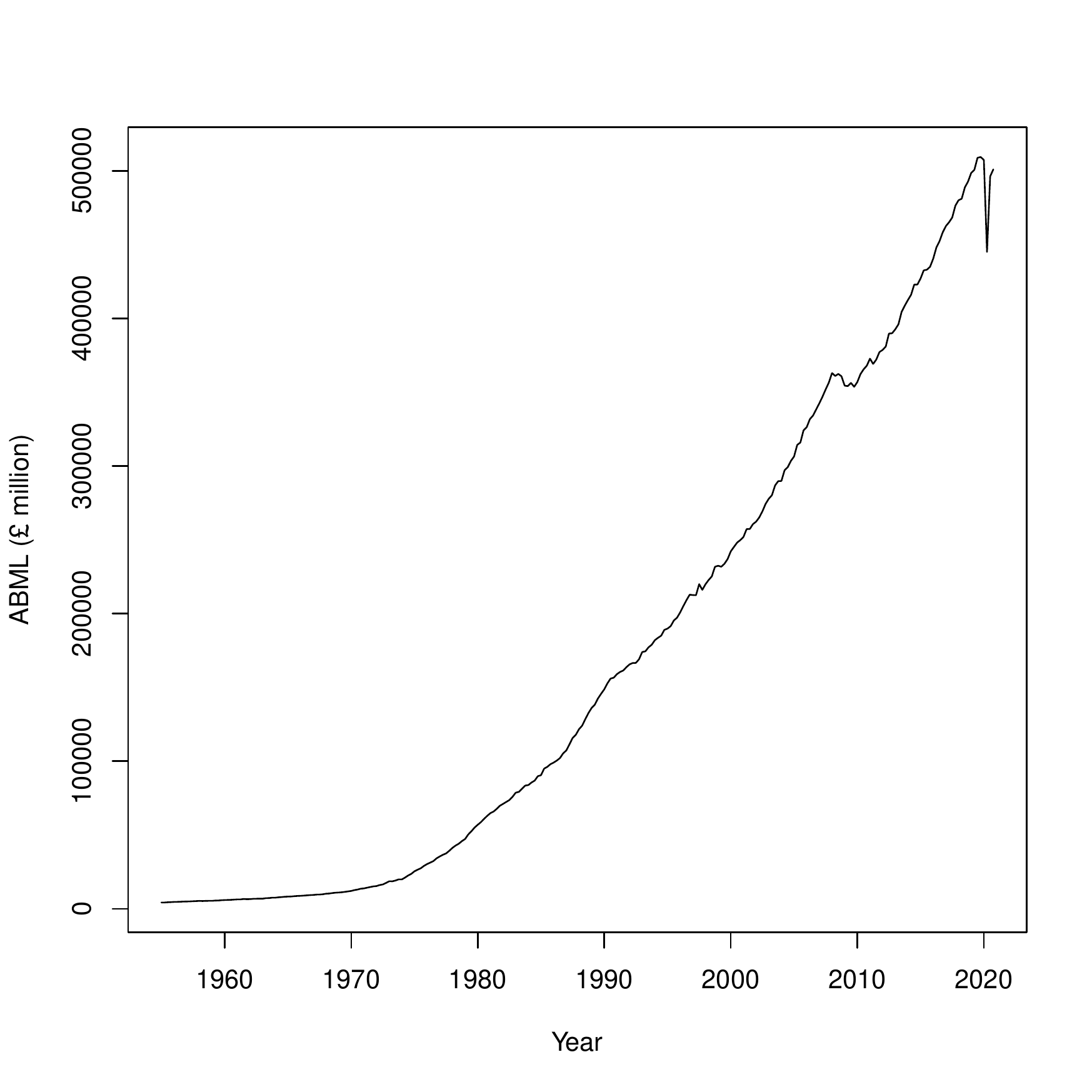}}
\caption{The ABML time series.\label{fig:abml}}
\end{figure}

Our work is motivated by a problem arising from the accurate forecasting of economic time series.  Specifically we consider the ABML time series, which contains values of the U.K.\ gross value added (GVA), a major component of the U.K.\ gross
domestic product (GDP). Both are vitally important economic statistics, with accurate forecasts being crucial in measuring the size of and growth in the UK economy.  Our ABML series is recorded quarterly from quarter one 1955 until quarter four 2020, consists
of $T=264$ observations and is plotted in Figure~\ref{fig:abml}. The impact of the `great financial crisis'
of 2008 and the COVID pandemic during 2020 can be clearly seen.
The data can be acquired from the Office for National Statistics website
{\tt https://www.ons.gov.uk}.

\begin{figure}[!h]
\centering
\subfigure[]{\includegraphics[width=0.49\textwidth]{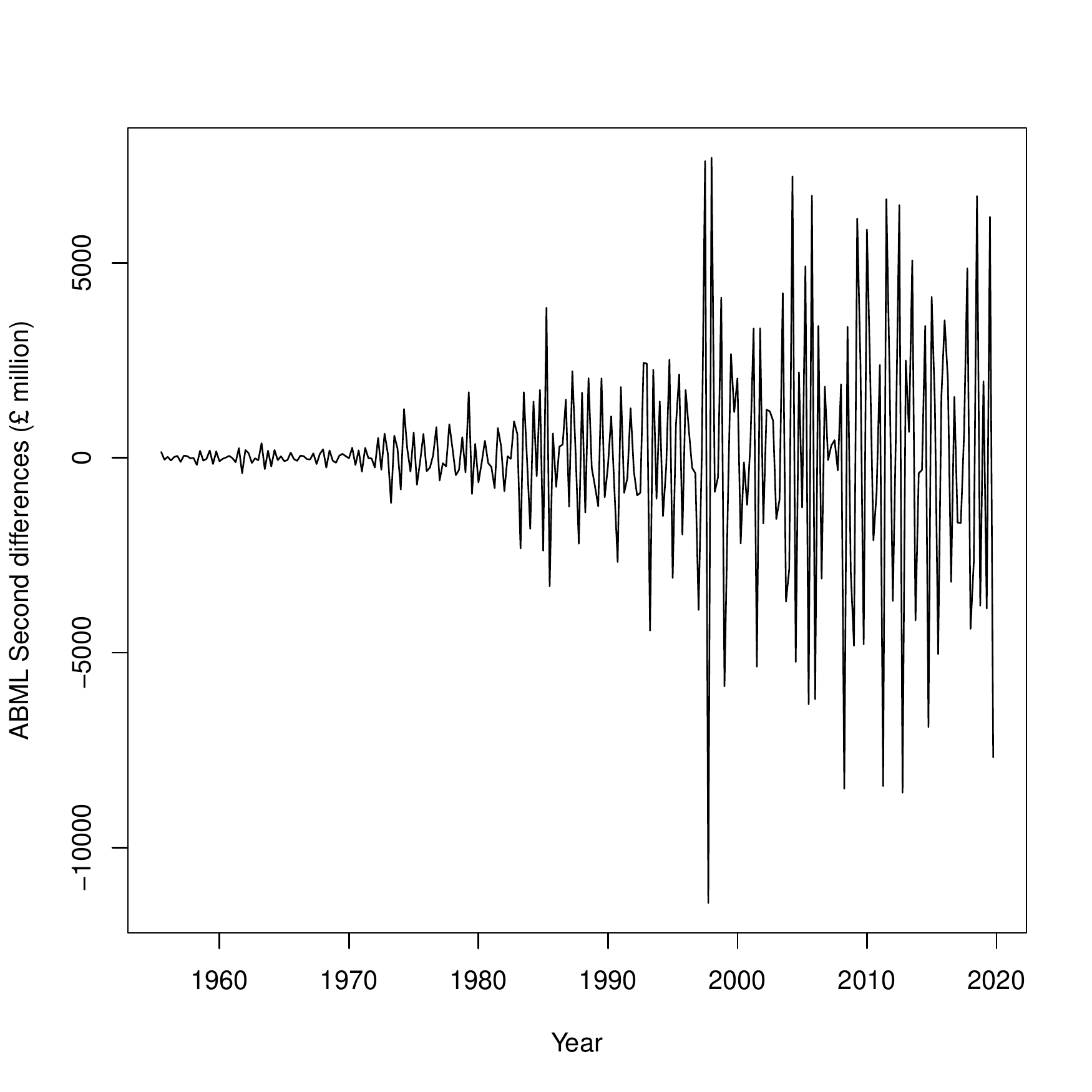}} \label{fig:abml2}
\subfigure[]{\includegraphics[width=0.49\textwidth]{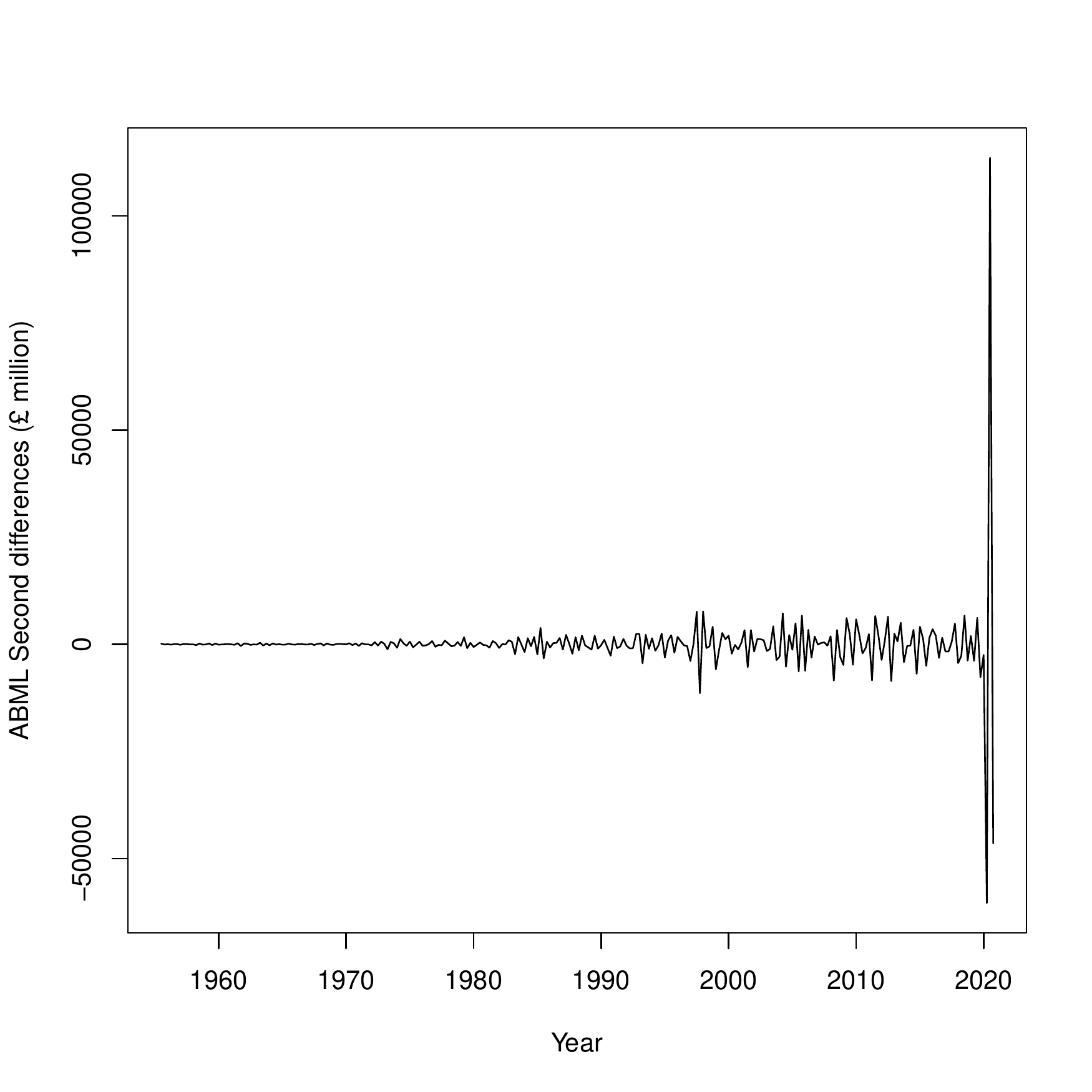}} \label{fig:abml3}
\caption{The second differences of the ABML time series (a) with and (b) without the COVID pandemic period.}\label{fig:abmlA}
\end{figure}

As with many economic time series, ABML exhibits a clear polynomial-like trend, which is characteristic of an
integrated economic time series.  Using standard statistical time series procedure, e.g.\ \cite{chatfield03:the},
we remove the trend
using second-order differences.
The second differences of our ABML series, including and not including the COVID period (up to Q4 2019),
are shown in 
%Figures~\ref{fig:abml2} and~\ref{fig:abml3}
Figure~\ref{fig:abmlA}, the latter amply demonstrating
the dramatic impact of the COVID pandemic.
\begin{comment}
\begin{figure}[!h]
\centering
\resizebox{\mywidth}{!}{\includegraphics{abmldiff21}}
\caption{The second differences of the ABML time series without the COVID period.\label{fig:abml2}}
\end{figure}
\begin{figure}[!h]
\centering
\resizebox{\mywidth}{!}{\includegraphics{abmldiff21_inCOVID}}
\caption{The second differences of the ABML time series including the COVID period.\label{fig:abml3}}
\end{figure}
\end{comment}
 
Both figures strongly suggest that the series is not second-order stationary, in that the variance of the series increases markedly over time.
Methods from~\cite{nason13} show that
 the correlation also changes over time.
Much of the increase in variance observed in Figure~\ref{fig:abml} is probably due to inflation. However, we  also analysed two different inflation-corrected versions of ABML, one provided by the U.K. Office of National Statistics, and both of these reject the null hypothesis of
second-order stationary, as determined by tests of
stationarity in \cite{PriestleySubbaRao1969} and \cite{nason13}.
Consequently, given that our series is nonstationary, to attempt forecasting we ought to use methods designed for such series.

The fundamentals of nonstationary forecasting have been considered by several authors, but mostly from a theoretical
standpoint. See, for example, \cite{Whittle63}, \cite{AP67}, \cite{SubbaRao73}
and \cite{Hallin86}. \cite{dahlhaus96:on} uses a version of Kolmogorov's formula
\citep[Theorem 5.8.1]{brockwell91:time} for zero-mean locally stationary time series, paralleling the \cite{SubbaRao73} result for the oscillatory process model.  Several other methods for nonstationary series forecasting exist, such as the neural network method of \cite{chow96:neuralnet}, \cite{ms04} or the simple, but effective, time-varying unconditional variance model of \cite{vbvs04}.

 \cite{fryz03:forecasting} derived time-varying Yule-Walker equations for the locally stationary wavelet (LSW) model of \cite{nason00:wavelet},
introducing local Yule-Walker estimators and solvers.  The authors carefully established the theoretical properties of their estimators and proposed a practical forecasting methodology, which included
parameter selection advice. % and working {\tt S} software (available from the first author).
However, as a grid search based approach, it can be computationally intensive to implement in practice. 

Our aim in this article is to develop a new, readily implemented, automated approach to forecasting LSW time series.  Here, we assume future observations can be modelled in terms of  present and recent past values and adopt the stance of ~\cite{fryz03:forecasting}, thus restricting our attention to predictors that are linear functions of the data and seeking to minimize their associated mean square prediction error. We use the recently proposed local partial autocorrelation \citep{killick19:lpacf} to dynamically select
the number of past time series observations to use in forecasting (denoted by $p$ below), rather than relying on the grid-type search proposed by~\cite{fryz03:forecasting}. Our approach is both computationally simpler and provides much improved forecasting results, which we demonstrate on both simulated and real data.

Section~\ref{sec:review} details the
LSW forecasting framework from~\cite{fryz03:forecasting}.
Section~\ref{sec:lpacfforecasting} introduces our improvement, uses of the local partial autocorrelation function to estimate
the local structure required for forecasting.  Section~\ref{sec:simulations} demonstrates this improved performance on a range of
simulated stationary and nonstationary time series.
Such a simulation study is required as although we strongly suspect our ABML gross value
added series is nonstationary, we cannot be absolutely sure. Hence, we need at least some validation of our new methodology
on both stationary and nonstationary series.
Section~\ref{example:abml} provides our forecasting results on the ABML
economic data series.
Section \ref{sec:discussion} provides further discussion and concludes.

\section{Review of Locally Stationary Wavelet Time Series Forecasting}\label{sec:review}

We begin by reviewing the locally stationary wavelet (LSW) model \citep{nason00:wavelet} and its associated forecasting framework \citep{fryz03:forecasting}. The LSW approach for modelling nonstationary time series has been used in many fields, from climatology \citep{Fryzlewicz03} and ocean engineering \citep{killick2013a} to biology \citep{hargreaves19:wavelet}, medicine \citep{nason2015bayesian, embleton22:multiscale} and finance \citep{fryzlewicz2005modelling}.

\subsection{The Locally Stationary Wavelet Model}\label{sec:lsw}

The LSW model \citep{nason00:wavelet, Fryzlewicz03} encompasses sequences of doubly-indexed stochastic processes
$\{ X_{t,T} \}_{t=0, \ldots, T-1}$, $T = 2^J \geq 1$, having the following representation in the mean-square sense
\begin{equation}
\label{eq:lswdef}
X_{t, T} = \sum_{j=1}^J \sum_k w_{j,k; T}  \psi_{j,k}(t) \xi_{j, k},
\end{equation}
where $\xi_{j,k}$ is a random orthonormal increment sequence and the $\{ \psi_{j, k} (t) \}_{j, k}$
form a discrete non-decimated family of wavelets based on a mother wavelet, $\psi(t)$, of compact support.
The quantities in~\eqref{eq:lswdef} are assumed to observe a number of key properties. Most notably
$\E (\xi_{j,k}) = 0$ for all $j, k$, and hence $E(X_{t, T}) = 0$ for all $t, T$, together with an assumption that
$\cov{\xi_{j, k}, \xi_{\ell, m}} = \delta_{j, \ell} \delta_{k, m}$ where $\delta_{i,j}$ is the Kronecker delta.  In addition,
\cite{nason00:wavelet} also introduces a number of conditions to ensure the amplitudes $\{w_{j,k;T}\}_k$ vary slowly within each level, thereby controlling the degree of local stationarity of the process.

Within the LSW framework the evolutionary wavelet spectrum (EWS), defined as $S_j(z) = |W_j (z)|^2,$ at each scale $j = 1, \ldots, J$ and rescaled time $z=k/T \in (0,1)$, plays an analogous role to that of spectrum in the stationary time series setting. The EWS quantifies the process power distribution over time and scale, and is connected to a localised autocovariance function defined for each (rescaled) time $z$ and lag $\tau\in\mathbb{Z}$ as follows
\beq\label{eq:lacv}c(z,\tau)=\sum_{j=1}^\infty S_j(z)\Psi_j(\tau).\eeq
The  $\{\Psi_j(\tau)\}_j$ is a family of compactly supported autocorrelation wavelets,
see \citep{nason00:wavelet}.

Spectral estimation is usually carried out by means of the raw wavelet periodogram, defined as $I_{j,k;T} = \lvert d_{j,k;T} \rvert^2$, where $d_{j,k;T} = \sum_{t=0}^T X_{t, T} \psi_{j,k}(t)$ are the empirical nondecimated wavelet coefficients. For notational simplicity, we shall refer to the raw periodogram as $I_{j,k}$.

An asymptotically unbiased estimator of the EWS is provided by the (corrected) empirical wavelet spectrum 
\begin{equation}
\label{eq:corrected per}
 \mathbf{L}(z) = A^{-1} \mathbf{I}(z),
\end{equation}
for all  $z \in (0,1)$, where $\mathbf{I}(z):= (I_{j,[zT]})_{j=1}^J$  is the raw wavelet periodogram vector and $A$ is a $J \times J$ symmetric matrix with entries $A_{j,\ell}=\sum_\tau \Psi_j(\tau)\Psi_\ell(\tau)$. As in the stationary setting, the wavelet periodogram is not a consistent estimator of the wavelet spectrum \citep{nason00:wavelet}. One way to overcome this is to smooth the raw wavelet periodogram as a function of (rescaled) time within each scale $j$, and then to apply the correction by $A^{-1}$ as in~\eqref{eq:corrected per}. %In what follows, we denote the corrected and smoothed periodogram of a time series $\{X_{t,T}\}_{t=0}^{T-1}$ as $\{\hat{S}_{j}(z)\}_{j}$, for rescaled time $z \in (0,1)$.

Once a well-behaved spectral estimator, $\hat{\mathbf{L}}$, has been obtained, equation~\eqref{eq:lacv} can be used to obtain a local autocovariance estimator $\hat{c}(z,\tau)$, e.g. in the notation above $\hat c(z,\tau)=\sum_{j=1}^J \hat{L}_j(z)\Psi_j(\tau).$
% and its approximate confidence bounds, under the assumption of Gaussian innovations \citep{nason2013}.

The LSW framework has proved to be useful across a variety of tasks when compared to competitor methods, e.g. for classification \citep{fryz09:consistent, krzemieniewska2014classification},
clustering \citep{hargreaves17:clustering}, testing for stationarity \citep{nason13}, spectral equality \citep{hargreaves19:wavelet} and replicate-effect \citep{embleton22:wavelet}, changepoint detection \citep{Nam15:LSWcptUncertainty} and testing for white noise and aliasing \citep{eckley18:aliasing}. We focus here on  forecasting nonstationary time series, and in particular on the work of \cite{fryz03:forecasting}.

\subsection{Forecasting within the LSW framework}            	
\subsubsection{Existing work} 
Given observations $X_{0,T},\ldots,X_{t-1,T}$ of a zero-mean locally stationary wavelet process, the method of \cite{fryz03:forecasting} ($\fryz$ algorithm henceforth) proposes to predict the next observation $X_{t,T}$ by taking a linear combination of the most recent $p$ observations
\beq\label{predeq}
\hat{X}_{t,T}=\sum_{s=t-p}^{t-1}b_{t-1-s,T}X_{s,T}.
\eeq
Here the predictor coefficients $\bm{b}=(b_{0,T},\ldots,b_{p-1,T})$ are chosen to minimize the mean square prediction error (MSPE). This is similar to the typical forecasting approach in the stationary context, except the weights $\bm{b}$ depend on time.
Intuitively, the reason for not using the whole series when predicting the next
observation lies in the nonstationary character of the process: the beginning of the series might have a different structure to the end and hence may not be useful for forecasting.

\cite{fryz03:forecasting} showed that
$\mbox{MSPE}(\hat{X}_{t,T},X_{t,T})=\mathbb{E}(\hat{X}_{t,T}-X_{t,T})^2$
can be approximated by
$\bm{b}^T B_t \bm{b}$, where $B_t$ is a $(p+1) \times (p+1)$ matrix whose $(m,n)$th entry is given by
%%\beq\nonumber
$(B_t)_{m,n}=\!\sum_{j=1}^J S_{j}\!\left(\frac{n+m}{2T}\right)\Psi_{j}(m-n)=c\left(\frac{m+n}{2T},m-n\right).$
%%\eeq
The weights $\bm{b}$ can then be obtained by solving the `generalised' set of Yule-Walker
equations
\beq\label{eq:yule}
\sum_{s=t-p}^{t-1}b_{t-1-s,T} \ c\left(\frac{n+s}{2T},s-n\right) =
c\left(\frac{n+t}{2T},t-n\right), \forall n=t-p,\ldots,t-1,
\eeq
which can then be extended to \mbox{$h$-steps-ahead} prediction.  These weights may be estimated by plugging in a good estimator of the local autocovariance function, with the forecast quality obviously highly reliant on the quality of these estimators.

Contrasting the usual approach in LSW estimation discussed in Section ~\ref{sec:lsw}, the $\fryz$ method proposes to obtain an estimate of the wavelet spectrum $\{\hat{S}_{j}(k/T)\}_{j}$ at all rescaled times corresponding to observed times (up to $(t-1)$) without smoothing and then to obtain an estimated local autocovariance $\tilde{c}(k/T,\tau)$ at each time $k={t-p,\ldots,t-1}$ and lags $\tau$. For consistency, \cite{fryz03:forecasting}  smooth the estimated local autocovariance by means of a standard kernel smoothing method  (using a normal or box kernel) with a chosen bandwidth that the authors denote by $g$. As well as providing a consistent estimator for the observed times, smoothing the local autocovariance additionally allows the forward estimation of the local autocovariance at rescaled time $t/T$. This (smoothed) estimated local autocovariance can then be used in the generalised Yule-Walker equations~\eqref{eq:yule} and the weights $\hat{\bm{b}}$ derived. The forecast
$\hat{X}_{t,T}=\sum_{s=t-p}^{t-1} \hat{b}_{t-1-s,T} X_{s,T}$  is then obtained, as well as its associated prediction error. \cite{fryz03:forecasting} generalised the above one-step-ahead prediction to an $h$-steps ahead.

\subsubsection{Criticisms of the $\fryz$ Method}\label{sec:regularize}
Whilst theoretically tractable, \cite{xie09:forecasting} noticed that the $\fryz$ algorithm sometimes produces abnormally large forecasts.
They identified the cause to be the occasional near-singularity of local covariance matrices and proposed a modification of the algorithm in order to stabilise the forecasts.
Their proposal constrains the Yule-Walker solution vector $\bm{b}$ to have unit norm.
The authors note that the revised method consistently produces better forecasts for a variety of prediction horizons.

A key quantity in calculating the forecast in equation~\eqref{predeq} is $p$: the amount of recent
data used for prediction.
\cite{fryz03:forecasting} and \cite{xie09:forecasting} used an adaptive grid search method to select $p$, which requires a starting value $p_0$.  In their approach, from one time point to the next, $p$ can only increase or decrease by one thus
sometimes resulting in a slow adaption to the dynamics of the evolving series. 

$\fryz$ suggested a procedure, also adopted by \cite{xie09:forecasting}, for simultaneously selecting the parameters $p$ (the number of most recent observations from the past used to inform the forecast $\hat{X}_{t,T}$) and $g$ (the bandwidth of the smoothing kernel). Briefly, their algorithm starts with some initial values for parameters $p$ and $g$, and the pair $(p,g)$ then are updated in an iterative process that evaluates their corresponding prediction performance for known data. The underlying idea is that the $(p,g)$ pair gets trained over a segment of length $m$ at the end of the series. \cite{fryz03:forecasting} proposed to choose $m$ to be the length of the longest segment at the end of the observed series with an apparent stationary behaviour, judged by {\em visual inspection}. Having made a choice for the parameter $m$, the $\fryz$ algorithm can be summarised as follows
\begin{enumerate}
\item Make an initial choice of parameters, say $(p_0,g_0)$, and use it to obtain predicted $\hat{X}_{t-m,T}$ by means of equation~\eqref{predeq} using the previous $p_0$ process observations.
\item Predict ${X}_{t-m,T}$ also by using pairs of parameters around $(p_0,g_0)$, \i.e. $(p_{0} \pm 1,g_{0}\pm{\delta})$ for some fixed constant $\delta$.
\item The pair that gives `closest' forecasts (in the sense of the minimum relative absolute prediction error) to the observed process is chosen, $(p_1,g_1)$, say.
\item Repeat steps (a)--(c). The updated pair of parameters is then itself updated through predicting $X_{t-m+1,T}$ from the previous $p_1$ observations and, by re-iterating this process, a parameter pair
$(p,g)$ is obtained for predicting the desired $X_{t,T}$.
\end{enumerate}
Hence, despite this `automatic' tuning, the practitioner must decide on the initial parameter pair $(p_0,g_0)$, the training length $m$ and the smoothing kernel (normal or box) used in the $\fryz$ method.

\section{Automating the Locally Stationary Wavelet Forecast} \label{sec:lpacfforecasting}

Our proposal departs from the currently adopted practice by  automating the forecasting procedure. Instead of the smoothing and forward estimation being undertaken at the level of the local autocovariance function, we take advantage of the recent advances in local spectral estimation \citep{nason13} and propose to perform running mean {\em periodogram} smoothing, from which forward estimation followed by correction are straightforward.  This avenue affords local autocovariance estimation through equation~\eqref{eq:lacv} and yields estimators with desirable properties \citep{nason13}. In turn, this approach also removes the need for the initial bandwidth choice ($g_0$) and training process since our proposed periodogram smoothing contains an automatic bandwidth selection.  In addition, we propose to remove the choice of $p_0$ and segment length $m$ by adopting a {\em localised} estimator for $p$ in order to determine the (time-dependent) forecasting window. Specifically, we  use the
local partial autocorrelation function (lpacf)  proposed by \cite{killick19:lpacf}, as a measure of the localised conditional correlation structure. Details of this approach are provided below. 

\subsection{Proposed lpacf-based forecasting ($\us$)}

Intuitively, the choice of $p$ amounts to establishing the length of a (sub)interval over which the process displays stationary behaviour, so that data over this interval can feasibly contribute to the linear prediction. This is evocative of the stationary autoregressive setting where the unknown dependence order $p$ is chosen via the partial autocorrelation function, $q$, using the theoretical property $p = \min\{\tau : q(\tau) = 0\}$ \citep[p.~36]{tsay02:analysis}. Of course, in practice, estimates $\hat{q}(\tau)$ are computed and, as these will never be exactly zero, their associated confidence intervals can be used to obtain an estimator $\hat{p}$ (see e.g. Theorem 8.1.2
from \cite{brockwell91:time}). This method of selecting the number of past observations to feed into prediction is widely used, even when the assumed underlying process is not autoregressive. 

Our proposal is to use a similar approach in which a {\em localised version of $p$} (and its estimate) is used to inform the length of the forecasting window.  In our approach we adopt the recently proposed local partial autocorrelation function (lpacf) of \cite{killick19:lpacf}, as well as a corresponding estimator. We next define their lpacf before demonstrating how we can apply it to nonstationary forecasting.

Mathematically, \cite{killick19:lpacf} define the local partial autocorrelation function $q(z,\tau)$ at rescaled time $z$ and lag $\tau$ as follows. 
\begin{definition}
\label{def:lpacf:lpacf}
	Let $\{X_{t,T}\}$ be a zero-mean locally stationary wavelet process with local autocovariance $c(z, \tau)$ and spectrum $\{ S_{j}(z)\}_j$ that satisfy
\begin{align}
\sum_{\tau = 0}^{\infty} \sup_z | c(z, \tau)| < \infty, \nonumber  %\label{eq:lswsm} \\
C_1 := \ess \inf_{z,\omega} \sum_{j > 0} S_j(z) | \hat{\psi}_j (\omega) |^2 > 0, \nonumber %\label{eq:posspec}
\end{align}
where $\hat{\psi}_j (\omega) = \sum_s \psi_{j, 0} (s) \exp(i \omega s)$.  Then, the local partial
autocorrelation function (lpacf) at (rescaled) time $z$ and lag $\tau$ is given by:
	\begin{equation}\label{deflpacf}
		\p{z}{\tau} =\phis \left\{ \frac{(\bb)^T\Bb\bb}{(\bf)^T\Bf\bf} \right\}^{1/2},
	\end{equation}
where
\begin{enumerate}
\item the quantity $\phis$ is the last element in the vector $\PPhis$ (of length $\tau$) obtained as
the solution to the local Yule-Walker equations i.e.\ $\BT \PPhis=\rhs_{[zT]}$,

\item the matrices $\Bf$ and $\Bb$ are  the local approximations of $\sigf$ and
$\sigb$, as  in the proof of Lemma A.1 from \cite{fryz03:forecasting}.

\item the coefficient vectors $\bf$ and $\bb$ are obtained as the solution to the
forecasting and back-casting prediction equations, or equivalently through minimisation
of the $\mmspe$. See Section~3.1 and Proposition~3.1 from \cite{fryz03:forecasting} for details.
\end{enumerate}
\end{definition}
The right hand term under the square root in \eqref{deflpacf} is a measure of nonstationarity.  For a stationary process, the square root term equals one and the localised $q(z,\tau)$ coincides with the classical partial autocorrelation measure $q(\tau)$. If there is a degree of nonstationarity within the data, then $q(z,\tau)$ will be modified by the nonstationarity factor.

\cite{killick19:lpacf} propose two methods for estimating the lpacf.  
For the purposes of forecasting we prefer to use the second, more stable windowed estimator,
denoted $\tilde{q}_W(z,\tau)$, which has been shown to have an asymptotically normal distribution and practically to work well both in simulated and real data settings \citep{killick19:lpacf}. 
Crucially, this windowed estimator allows for the {\em local} estimation of the partial autocorrelation at the last observation in the process ($t-1$), as this is the point around which the prediction is made. Corollary 1 from \cite{killick19:lpacf} then allows us to construct confidence bounds for the local partial autocorrelation function. Recall that choosing $p$ in the stationary setting is akin to estimating the number of significant lags in the partial autocorrelation function. We mimic this idea and obtain the estimate $\hat{p}$ as the largest significant lag $\tau$ in the confidence interval of $\tilde{q}_W(z,\tau)$ at rescaled time $z=(t-1)/T$.  However we stress that we do not necessarily assume that the underlying process is necessarily autoregressive of any order. Future work could investigate whether $\hat{p}$ is a consistent estimator of the true value of $p$ {\em if} the underlying process was indeed locally autoregressive.

Algorithm \ref{alg:forlapalgo} details the steps of our proposed forecasting algorithm, $\us$. Whilst our theoretical framework is based on~\cite{fryz03:forecasting}, our
practical implementation differs considerably.  Specifically, in addition to (i) using
$\hat{p}$ as an appropriate value of $p$, we also (ii) use the recent locally stationary wavelet
covariance and bandwidth estimation from
\cite{nason13}, implemented in the \texttt{locits} {\tt R} package \citep{R:locits}, and
(iii) permit the covariance matrix regularization method of
\cite{xie09:forecasting} as an option.

%%%algorithm%%%

%%%%%%%%%%%%%%%%%%%%%%%%%%%%
% temporarily redefine the figurename for this float only.
\renewcommand{\figurename}{Algorithm}
\newcounter{oldcnt}
\setcounter{oldcnt}{\value{figure}}
\setcounter{figure}{\value{algno}}

\begin{figure}[!h]
\begin{algorithm}{lpacf-based forecasting algorithm ($\us$)}\\
Assume we observed $\{X_{0,T}, \dots, X_{t-1,T}\}$ with $T = 2^J$.
\begin{enumerate}

\item  \textit{Determine $p$ via lpacf estimation}: obtain the lpacf estimate $\tilde{q}_W$ \citep{R:lpacf} corresponding to time $(t-1)$ and set $\hat p$ to be the largest significant lag in its confidence interval.

\item  \textit{Spectral estimation:} estimate the spectral content of the observed signal by correcting (with the matrix $A^{-1}$, see equation~\eqref{eq:corrected per}) the running mean smoothed raw periodogram $\tilde{I}_{j,k}=(2s+1)^{-1}\sum_{u=k-s}^{k+s}I_{j,u}$, where the bandwidth $s$ is obtained automatically as described in \cite{nason13, R:locits}. Note that this procedure also embeds forward smoothing of the raw periodogram and thus enables the estimation of $\hat{S}_j(k/T)$ for $k=0, \ldots, t$.

\item  \textit{Autocovariance estimation:}    Estimate the local autocovariance $c(k/T,\tau)$ by means of equation~\eqref{eq:lacv} at rescaled times $z$ corresponding to observed times up to $(t-1)$ and lags $\tau$ as dictated by the generalised Yule-Walker equations~\eqref{eq:yule}. Also obtain $\hat c(t/T,\tau)$ by making use of the extrapolated spectrum  $\hat{S}_{j}(t/T)$.

\item \textit{Solve the generalised Yule-Walker equations}: obtain the estimated time-dependent weight vectors $\hat{\bm{b}}$ by solving equations~\eqref{eq:yule} over the most recent $\hat p$ observations, subject to the regularisation constraint of \cite{xie09:forecasting} if desired.

\item \textit{Forecast $X_{t,T}$}: by using the linear combination of the last $\hat p$ observations with weights $\hat{\bm{b}}$, as described in equation~\eqref{predeq}. The $1-\alpha$ {\em prediction interval} uses the corresponding estimated MSPE: $\hat{X}_{t,T}\pm  \Phi^{-1}(1-\alpha/2) \left(\hat{\bm{b}}^T \hat{B_t} \hat{\bm{b}}\right)^{1/2}$.
\end{enumerate}
\end{algorithm}
%\vspace{-1cm}
\caption{Proposed $\us$ algorithm for nonstationary time series forecasting. \label{alg:forlapalgo}}
\end{figure}
\renewcommand{\figurename}{Figure}		%return the figurename back to original
\setcounter{figure}{\value{oldcnt}}				%return figure counter back to where it was.
\addtocounter{algno}{1}

%%%%%%%%%%%%%%%%%

\section{Simulation Study}\label{sec:simulations}

 We assess the performance $\us$ by simulating from a variety of scenarios, including both stationary and nonstationary examples, as detailed below.
Our overarching goal is to demonstrate the versatility of our proposed $\us$ forecasting technique and its utility in the `toolkit' of any data analyst interested in forecasting. Our test set includes locally stationary wavelet processes as well as processes not represented by this framework. In so doing, we aim to assess the forecast performance of our approach both on realisations of wavelet processes but also a variety of other important model classes.

Below, we compare $\us$ to
(i) the \cite{fryz03:forecasting} method (FVBvS),
 (ii) forecasting using time-varying autoregressive processes (TVAR) of order $2$ and higher,
(iii) the Box-Jenkins forecasting procedure (B-J), designed for stationary time series but commonly used by analysts even on nonstationary data (see e.g., the report from \cite{ucmga08}), as well as (iv) the often employed exponential smoothing (ES) \citep{hynd}. 

The simulations, and real data example in later sections,  use the {\tt forecast} package \citep{R:forecast}, {\tt tvReg} package \citep{R:tvreg} and the {\tt smooth} package in the {\tt R} statistical programming language \citep{Rcore}. Specifically, we use the {\tt auto.arima} function in the former package that automatically chooses and fits an ARIMA model and in addition its {\tt forecast} function for forecasting, while time-varying autoregressive forecasting is carried out using the {\tt forecast} function in the latter package. Our $\us$ algorithm is implemented within the {\tt forecastLSW} package \citep{R:forecastLSW},
also in {\tt R}. Forecasts are produced using the method described in Section~\ref{sec:lpacfforecasting} and implemented in the {\tt forecast.lpacf} function, which makes use of the lpacf function estimate $\tilde{q}_W$ implemented in the {\tt lpacf} {\tt R} package \citep{R:lpacf}.

In the simulations, we consider one-step-ahead forecasts over the stretch of last 20 observations, $h=1$, although other horizons can be used.
In addition, the methods and associated software implementation work for arbitrary length time series even though some example series below happen to be of dyadic length.

Our study presents empirically computed measures for the prediction interval coverage rates and accuracy for all forecasting methods. Nominal rates ranging from 40\% to 90\% in steps of 10\% are used. The accuracy measure we adopt is the interval score of \cite{gneiting12:intervalscore} which simultaneously penalises wide prediction intervals and lack of coverage. In this setup, lower interval scores correspond to better prediction intervals.

The results in Tables~\ref{tab:epicr},~\ref{tab:epicrns} and~\ref{tab:epicrns2} are based on averages taken over
$K=500$ runs. Specifically, the final columns of each table contain for each method (i) the mean of the prediction coverage ratios (MCR) and (ii) the mean interval score (MIS) at the 90\% level, both relative to Box-Jenkins.  An MCR value greater than one demonstrates that the respective method provides better predictive coverage than the classical Box-Jenkins, while MIS values less than one indicate that the respective method provides better prediction intervals, and balance between narrow range and coverage. We note that the tables only show results corresponding to the TVAR2 method, but similar results are also obtained when using higher orders (e.g. order 5), hence these are not reported here.

\subsection{Stationary series}
Table~\ref{tab:epicr} shows empirical prediction interval coverage rates for three
models each with $T=128$ and standard normal innovations.
Model~A simulates independent $N(0, 1)$ variates,
Model~B simulates from  a stationary AR$(1)$ model with
parameter $\alpha=0.7$,  and Model~C simulates from a
stationary MA$(1)$ model with parameter $\beta= -0.5$.

\begin{table}
 \caption{\label{tab:epicr}Empirical Prediction Interval Coverage Rates and Accuracy: stationary underlying series. MCR=Mean Prediction Coverage. MIS= Mean Interval Score. (Superior behaviour relative to B-J when $MCR>1$, $MIS<1$).}
 \fbox{%
 \begin{tabular}{cr|rrrrrr|rr}
 & & \multicolumn{6}{c|}{Nominal} & &\\
 &   &  40\% &  50\% &  60\% &  70\% &  80\% &  90\% & MCR & MIS \\\hline
&B-J & 39.3 & 48.5 & 58.5 & 68.6 & 79.0 & 89.2 &  &  \\
Model A&  $\us$ & 38.8 & 48.0 & 58.1 & 68.2 & 78.3 & 88.8 & 1.00 & 1.00 \\
$N(0,1)$&  FVBvS & 36.7 & 46.4 & 55.7 & 65.4 & 74.9 & 86.1 & 0.97 & 1.09 \\
&  TVAR2 & 39.1 & 48.4 & 58.5 & 68.8 & 78.9 & 89.2 & 1.00 & 0.99 \\
&  ES & 39.5 & 49.4& 59.4 & 69.3 & 79.5 & 89.7 & 1.01 & 1.00 \\
\hline
&B-J & 38.9 & 48.9 & 58.2 & 68.2 & 78.7 & 88.9 &  &  \\
Model B&  $\us$ & 36.8 & 46.0 & 54.9 & 65.0 & 75.1 & 85.5 & 0.96 & 1.38 \\
AR$(1)$&  FVBvS & 25.9 & 32.5 & 39.3 & 46.9 & 55.4 & 65.2 & 0.74 & 5.22 \\
&  TVAR2 & 29.4 & 37.2 & 45.3 & 53.4 & 62.8 & 74.7 & 0.84 & 1.66 \\
&  ES & 39.7 & 49.4 & 59.3 & 69.0 & 79.3 & 89.9 & 1.01 & 1.05 \\
\hline
&B-J & 39.2 & 49.3 & 59.1 & 69.1 & 79.8 & 89.4 &  &  \\
Model C&  $\us$ & 36.7 & 45.8 & 56.0 & 65.9 & 76.3 & 87.0 & 0.97 & 1.11 \\
MA$(1)$&  FVBvS & 32.2 & 41.1 & 49.6 & 58.8 & 68.3 & 79.3 & 0.89 & 1.49 \\
&  TVAR2 & 35.9 & 45.0 & 54.5 & 64.7 & 74.9 & 86.2 & 0.96 & 1.10 \\
&  ES & 40.1 & 50.2 & 60.2 & 70.4 & 80.7 & 90.6 & 1.02 & 1.09 \\
\hline
\end{tabular}
}

\end{table}

The prediction coverage rates and accuracy for Box-Jenkins are best across all models, with $\us$ and ES almost matching for coverage across all models. The accuracy of $\us$ and ES for models A and C is close to their nominal values (a very similar behaviour to TVAR2). For model B, ES has better accuracy, but still lower than that of B-J. While all methods do well for model A, FVBvS and TVAR are markedly inferior for model B, and to a lesser extent for FVBvS on model C.

As we designed this setup specifically to include stationary processes, {\em a priori} one would have expected Box-Jenkins to be significantly better as it is designed for stationary series. However, the results show that our proposed $\us$ method is competitive even though it is not designed for stationary data, while the other nonstationary methods underperform.

\subsection{Nonstationary series}
\label{sec:simnonstat}

\begin{table}
 \caption{Models D--H. Empirical Prediction Interval Coverage Rates and Accuracy: nonstationary underlying series. MCR=Mean Prediction Coverage. MIS= Mean Interval Score. (Superior behaviour relative to B-J when $MCR>1$, $MIS<1$). \label{tab:epicrns}}
 \centering
\fbox{%
\begin{tabular}{cr|rrrrrr|rr}
 & & \multicolumn{6}{c|}{Nominal} & &\\
 &   &  40\% &  50\% &  60\% &  70\% &  80\% &  90\% & MCR & MIS \\\hline
&B-J & 31.4 & 39.6 & 47.8 & 57.1 & 67.5 & 79.1 &  &  \\
Model D&  $\us$ & 30.1 & 37.5 & 45.5 & 54.4 & 64.5 & 76.3 & 0.96 & 1.08 \\
TVAR(1)&  FVBvS & 21.6 & 27.3 & 33.3 & 39.7 & 47.6 & 57.7 & 0.73 & 3.88 \\
&  TVAR2 & 27.6 & 34.6 & 42.2 & 50.1 & 59.6 & 71.3 & 0.90 & 1.21 \\
&  ES & 29.8 & 37.8 & 46.1 & 54.9 & 64.9 & 77.4 & 0.98 & 1.13 \\
\hline
&B-J & 25.2 & 32.2 & 39.2 & 47.8 & 56.8 & 69.4 &  &  \\
Model E&  $\us$ & 19.4 & 24.6 & 30.2 & 36.0 & 43.1 & 51.9 & 0.75 & 2.49 \\
TVAR(1)&  FVBvS & 22.9 & 28.5 & 35.1 & 41.5 & 49.1 & 57.9 & 0.84 & 502.36 \\
&  TVAR2 & 19.2 & 24.6 & 29.8 & 36.1 & 44.3 & 54.3 & 0.77 & 1.92 \\
&  ES & 18.6 & 24.1 & 30.1 & 36.9 & 44.8 & 56.1 & 0.80 & 1.92 \\
\hline
&B-J & 29.8 & 38.2 & 46.6 & 55.4 & 65.7 & 77.80 &  &  \\
Model F&  $\us$ & 35.4 & 44.8 & 53.9 & 63.9 & 74.1 & 84.4 & 1.09 & 0.84 \\
TVAR(2)&  FVBvS & 29.5 & 37.4 & 46.3 & 55.0 & 64.8 & 76.0 & 0.98 & 1.45 \\
&  TVAR2 & 29.8 & 37.8 & 46.2& 55.3 & 65.4 & 76.9 & 0.99 & 0.93 \\
&  ES & 47.4 & 58.2 & 68.8 & 78.2 & 86.6 & 93.6 & 1.22 & 0.82 \\
\hline
&B-J & 36.6 & 45.8 & 55.5 & 65.0 & 75.4 & 86.1 &  &  \\
Model G&  $\us$ & 35.8 & 44.8 & 54.3 & 64.0 & 74.4 & 85.4 & 0.99 & 1.01 \\
TVAR(12)&  FVBvS & 33.3 & 41.7 & 50.2 & 59.5 & 69.4 & 80.5 & 0.94 & 1.32 \\
&  TVAR2 & 35.10 & 44.1 & 52.9 & 62.8 & 73.3 & 84.4 & 0.98 & 1.00 \\
&  ES & 37.8 & 47.3 & 57.3 & 67.1 & 77.5 & 88.1 & 1.03 & 0.96 \\
\hline
&B-J & 27.3 & 34.7 & 42.0 & 50.4 & 59.1 & 69.9 &  &  \\
Model H&  $\us$ & 33.1 & 42.4 & 50.9 & 60.1 & 69.4 & 80.2 & 1.17 & 0.66 \\
TVMA(1)&  FVBvS & 30.7 & 38.9 & 47.0 & 55.4 & 64.7 & 74.8 & 1.09 & 5.53 \\
&  TVAR2 & 32.5 & 42.3 & 49.8 & 59.4 & 71.0 & 82.6 & 1.20 & 0.53 \\
&  ES & 36.5 & 46.2 & 55.6 & 65.1 & 75.6 & 85.9 & 1.25 & 0.60 \\
\hline
\end{tabular}
}
\end{table}

\begin{table}
 \caption{Models I--M. Empirical Prediction Interval Coverage Rates and Accuracy: nonstationary underlying series. MCR=Mean Prediction Coverage. MIS= Mean Interval Score. (Superior behaviour relative to B-J when $MCR>1$, $MIS<1$).\label{tab:epicrns2}}
 \centering
\fbox{%
\begin{tabular}{cr|rrrrrr|rr}
 & & \multicolumn{6}{c|}{Nominal} & &\\
 &   &  40\% &  50\% &  60\% &  70\% &  80\% &  90\% & MCR & MIS \\\hline
&B-J & 33.1 & 41.9 & 50.7 & 60.2 & 70.6 & 82.3 &  &  \\
Model I&  $\us$ & 33.2 & 41.7 & 50.4 & 59.9 & 70.2 & 82.1 & 1.00 & 0.93 \\
TVMA(1)&  FVBvS & 28.9 & 36.2 & 43.8 & 52.0 & 61.4 & 73.0 & 0.89 & 1.66 \\
&  TVAR2 & 30.8 & 38.5 & 46.9 & 55.8 & 65.9 & 78.1& 0.95 & 1.01 \\
&  ES & 34.1 & 42.8 & 51.6 & 61.4 & 72.0 & 83.9 & 1.02 & 0.96 \\
\hline
&B-J & 22.2 & 28.2 & 34.7 & 41.5 & 49.7 & 60.6 &  &  \\
Model J&  $\us$ & 28.0 & 35.3 & 42.9 & 51.6 & 60.9 & 73.0 & 1.22 & 0.80 \\
TVMA(2)&  FVBvS & 31.4 & 39.4 & 48.0 & 56.9 & 67.0 & 78.2 & 1.32 & 0.88 \\
&  TVAR2 & 22.0 & 28.2 & 34.5 & 42.0 & 50.7 & 61.7 & 1.03 & 0.96 \\
&  ES & 22.6 & 29.1 & 35.4 & 43.0 & 51.6 & 62.9 & 1.05 & 0.96 \\
\hline
&B-J & 20.6 & 26.0 & 31.9 & 38.9 & 47.3 & 58.3 &  &  \\
Model K&  $\us$ & 26.3 & 33.4 & 40.5 & 48.7 & 58.2 & 70.4 & 1.23 & 0.79 \\
TVWN&  FVBvS & 29.8 & 37.9 & 46.3 & 55.4 & 65.6 & 77.0 & 1.35 & 1.00 \\
&  TVAR2 & 21.0 & 26.8 & 32.9 & 39.8 & 48.4 & 59.2 & 1.02 & 0.95 \\
&  ES & 21.3 & 27.0 & 33.2 & 40.1 & 48.8 & 59.7 & 1.03 & 0.96 \\
\hline
&B-J & 41.6 & 51.4 & 61.1 & 71.2 & 81.5 & 91.0 &  &  \\
Model L&  $\us$ & 41.9 & 52.3 & 62.5 & 72.6 & 82.4 & 91.7 & 1.01 & 1.05 \\
LSW-P3&  FVBvS & 31.3 & 39.1 & 47.0 & 56.0 & 65.9 & 77.4 & 0.85 & 4.03 \\
&  TVAR2 & 41.5 & 51.0 & 61.1 & 71.2 & 81.2 & 91.0 & 1.00 & 1.02 \\
&  ES & 43.3 & 53.8 & 64.4 & 74.7 & 84.5 & 93.4 & 1.03 & 1.02 \\
\hline
&B-J & 34.2 & 42.7 & 51.7 & 60.2 & 70.6 & 82.4 &  &  \\
Model M&  $\us$ & 36.8 & 45.8 & 55.5 & 65.1 & 75.1 & 85.4 & 1.05 & 1.02 \\
LSW-P4&  FVBvS & 33.7 & 42.1 & 50.9 & 59.7 & 69.7 & 80.8 & 0.99 & 2.24 \\
&  TVAR2 & 31.8& 40.0 & 48.6 & 57.9 & 67.6 & 79.3 & 0.97 & 1.02 \\
&  ES & 37.2 & 47.1 & 56.5 & 66.0 & 76.4 & 86.8 & 1.06 & 0.95 \\
\hline
\end{tabular}
}
\end{table}

{\em Model Specification.}
Tables~\ref{tab:epicrns} and~\ref{tab:epicrns2} show  coverage rates for ten nonstationary
models, summarised as follows. Models D and E are realisations of TVAR$(1)$
processes of different forms, whilst Models F and G are higher-order TVAR processes of orders two and
twelve, respectively. Models H--J correspond to different time-varying MA processes.
Model K is a
uniformly modulated white noise process~\citep[p.~826]{priestley82:spectral},
whilst models L and M are
locally stationary wavelet processes.
Complete model descriptions can be found in~\ref{app:sim.models}.
In each case the underlying innovations are independent
$N(0, \sigma^2_Z)$ random variables, and all
realizations are of
length $T=128$, except for model L where $T=512$, and model M where $T=350$ for variety.
\paragraph{Discussion of simulation results.} The $\us$ method provides much better coverage ($\mbox{MCR}\geq 1.05$) than Box-Jenkins for five of the models considered (F, H, J, K, M). For the models D, G, I, L the two methods
are broadly comparable. 
The method accuracy as measured by MIS provides a broadly similar picture, with $\us$ superior to B-J for five models F, H, I, J, K ($\mbox{MIS}\leq 0.95$), similar calibration for models G, L and M (MIS in the range 0.95--1.05) and lower for models D and E ($\mbox{MIS}\geq 1.05$).

For model E no approach proves particularly competitive. However, this is not altogether surprising as model~E is neither stationary nor, in fact, locally stationary with Lipschitz smoothness constraints as it experiences six changes in parameter, $\alpha(z)$, over only $T=128$ observations. Hence, a stationary forecasting system is clearly not appropriate and estimation by the nonstationary models is likely to be poor, as few observations contribute to any region of local stationarity. 

When compared to the Box-Jenkins method, the FVBvS method provides much better coverage ($\mbox{MCR}\geq 1.05$) for three of the models considered (H, J, K); of these, $\us$ outperforms FVBvS on model H (1.17 vs 1.09) and has lower performance on models J and K (1.22 vs 1.32, and 1.23 vs 1.35, respectively).  The accuracy of the FVBvS prediction intervals is lower ($\mbox{MIS}\geq 1.05$) than that of B-J for all models except for models J (superior) and K (the same).

When comparing TVAR coverage rates to those achieved by the Box-Jenkins baseline, TVAR outperforms B-J on model H and delivers comparable results (MCR values in the range 0.95--1.05) on all other models except for underperforming on models D and E. Unlike FVBvS, TVAR achieves similar levels of accuracy to B-J (MIS in the range 0.95--1.05) for models G, I, J, K , L, M, worse accuracy for models D, E, and better performance for models F and H.

The forecasting accuracy measure (MIS) consistently indicates that our proposed $\us$ technique delivers better calibrated prediction intervals than FVBvS on {\em all} models, with sizeable performance gaps in favour of $\us$. This metric indicates that  $\us$ and TVAR deliver comparably accurate prediction intervals for models G, L and M (ratio in the range 0.95--1.05), $\us$ is better for models D (1.08 vs 1.21), F (0.84 vs 0.93), I (0.93 vs 1.01), J (0.80 vs 0.96) and K (0.79 vs 0.95), and worse for models E (2.49 vs 1.92) and H (0.66 vs 0.53). According to the coverage rate measure, $\us$ is superior to TVAR on all models, except for models E, G, H, I and L where they deliver similar rates (corresponding to ratios in the range 0.95--1.05).

The simulations for model F are particularly revealing. The underlying model is
a TVAR$(2)$ and, indeed, the forecasting method based on that (TVAR2) performs somewhat
better (MCR 0.99, MIS 0.93) than the B-J method. However, it is interesting that our new method, $\us$, produces best coverage (1.09) and accuracy (MIS 0.84). Note that the other wavelet-based method, FVBvS, has comparable coverage to TVAR2 (0.98) at the price of sacrificing some accuracy (MIS 1.45).

In the TVAR2 setting, the real competitor for $\us$ turns out to be ES, which produces comparable results in terms of coverage and accuracy rates (corresponding ratios in the range 0.95--1.05) to $\us$ for all models, except for ES obtaining superior results on model H and $\us$ on models J and K, where ES performs closer to the B-J benchmark.

In addition to the simulations described above, we repeated our simulations on these models with \mbox{$t_4$-distributed}
innovations scaled to have unit variance for B-J, $\us$ and FVBvS.
With the heavier-tailed innovations the relative performance of Box-Jenkins compared to $\us$ forecasting changes negligibly: the largest MCR change is 0.05 and 70\% of the changes are less than 0.02.

Overall, one could conclude that, for zero-mean locally stationary series, the $\us$ method is much better than the Box-Jenkins method in two-thirds of cases and about the same in the remaining one-third, and provides an improvement over all of the other
nonstationary-based methods across the majority of our
models.

\section{Forecasting the U.K. National Accounts time series}\label{example:abml}

We now return to consider the ABML series that helped motivate this work in Section \ref{sec:data}. The Box-Jenkins method is sometimes used to forecast the ABML series,
see~\cite{ucmga08} for example. Table~\ref{tab:ones50} shows that FORLAP seriously
outperforms the Box-Jenkins methodology for forecasting the  second differences of
ABML series at one-step ahead. Here, the input data is provided to both methods up
to $t= T-n$ and $T-n+1$ is forecast and the percentages are average success rates
over $n=T-51, \ldots, T-1$.
\begin{table}
 \caption{Percentage of times that the Box-Jenkins and the FORLAP methods' 95\%
 	one-step ahead forecast prediction intervals contain the truth over the last
	50 time points. \label{tab:ones50}}
 \centering
 \fbox{%
\begin{tabular}{rrr}
			& \multicolumn{2}{c}{ABML series}\\
  	Method & With COVID & Without COVID\\\hline
Box-Jenkins	&    66 & 72\\
FORLAP		&  90   & 90\\\hline
\end{tabular}
}
\end{table}

\begin{figure}[!h]
\centering
\resizebox{\mywidth}{!}{\includegraphics{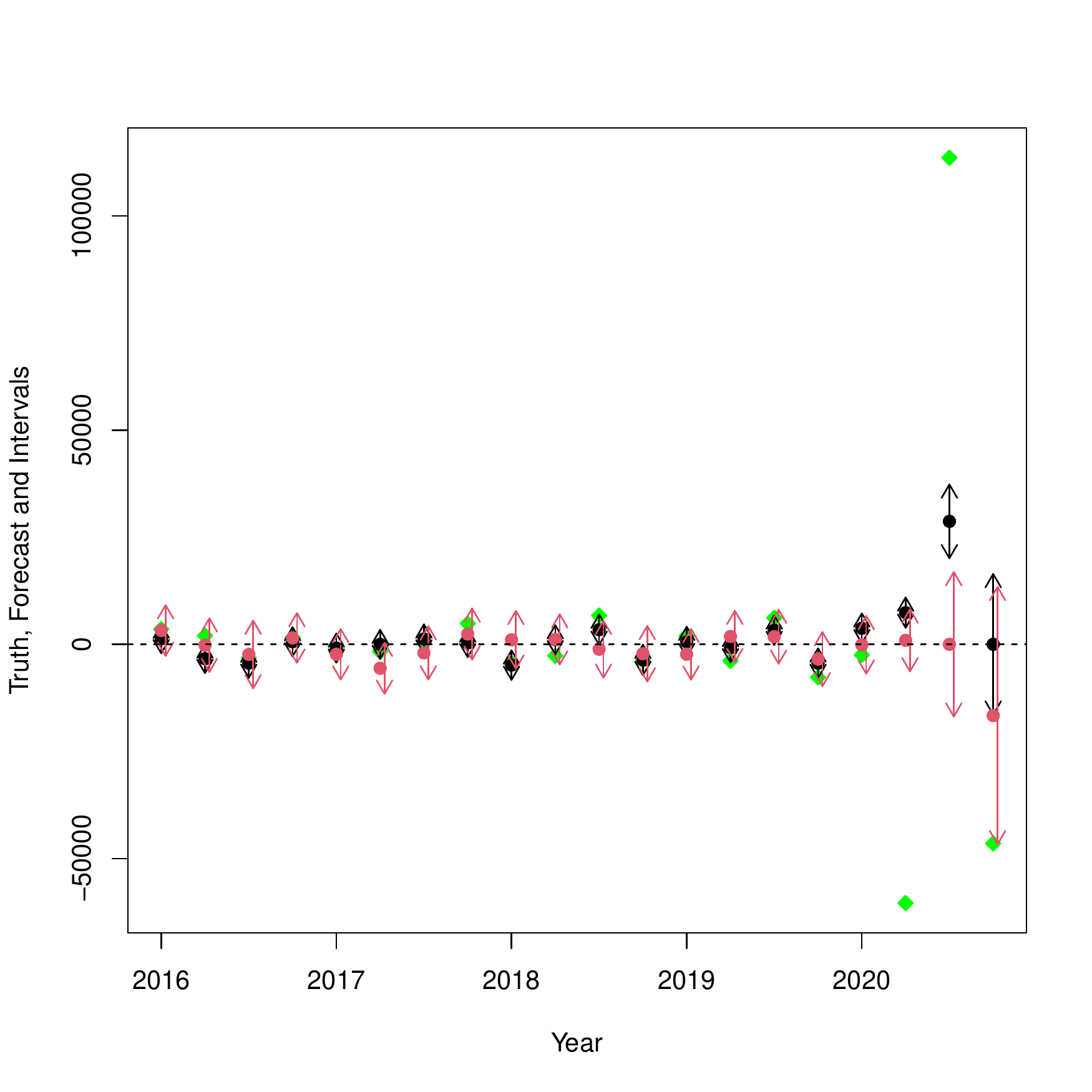}}
\caption{Forecast results for one-step ahead forecasting of the final twenty observations of the ABML
 series including the COVID period. Green diamond=True value of series. Black=Box-Jenkins. Red=$\us$.
 Solid circle indicates point forecast and double-headed
	arrows indicate
	95\% prediction intervals. \label{fig:abmlForIncCovid}}
\end{figure}

\begin{figure}[!h]
\centering
\resizebox{\mywidth}{!}{\includegraphics{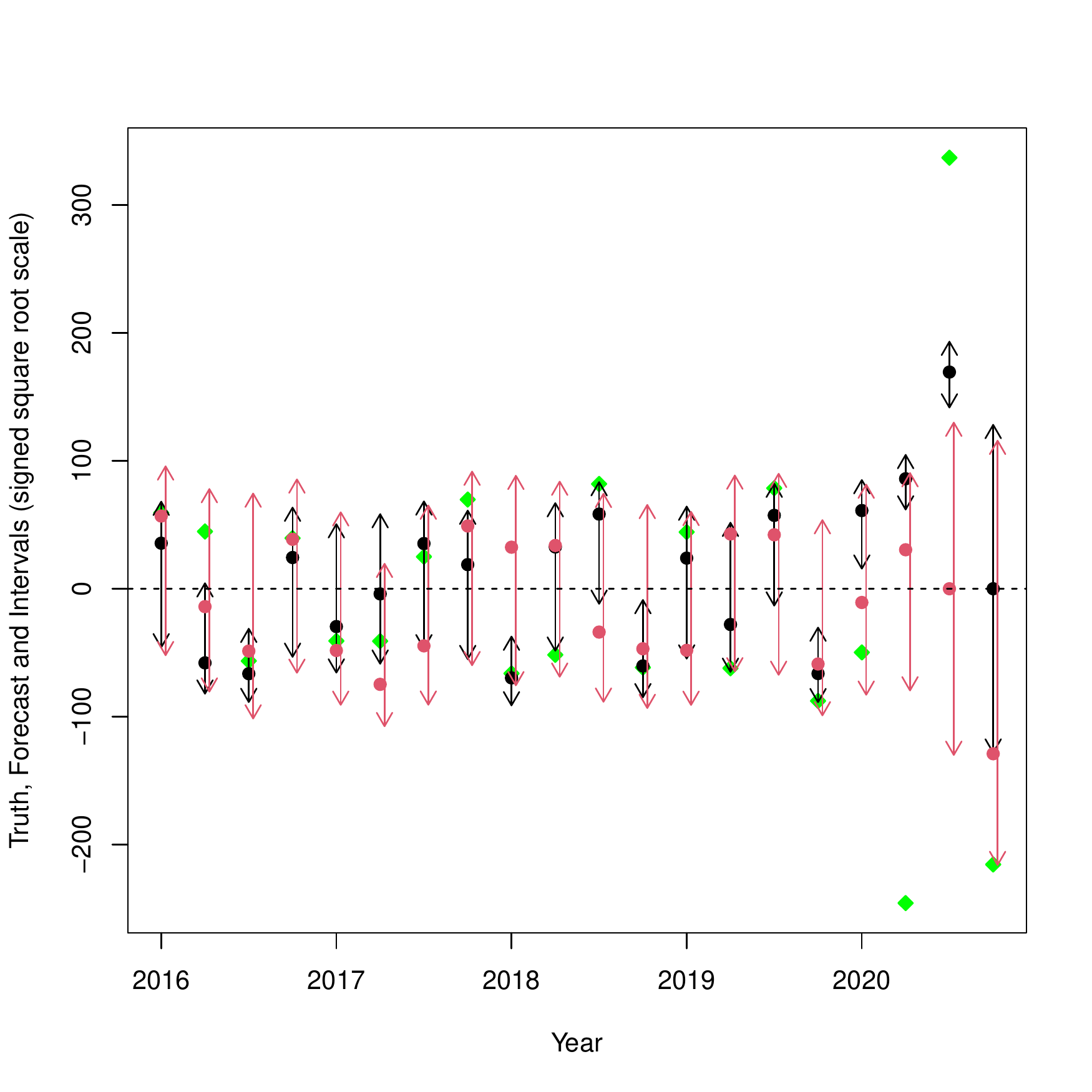}}
\caption{Forecast results for one-step ahead forecasting of the final twenty observations of the ABML
 series including the COVID period plotted on a signed square root scale.
 Green diamond=True value of series. Black=Box-Jenkins. Red=$\us$.
 Solid circle indicates point forecast and double-headed
	arrows indicate
	95\% prediction intervals. \label{fig:abmlForIncCovidT}}
\end{figure}

Figure~\ref{fig:abmlForIncCovid} compares $\us$ against Box-Jenkins forecasting for ABML in
a similar way to Table~\ref{tab:ones50} except for the last 20 time points.
The good performance of FORLAP relative to Box-Jenkins can be discerned, e.g.\ particularly Q1 and Q4
of 2020, but the sheer size of the movements during the pandemic obscures what the forecasts
are doing earlier in the plot. So, Figure~\ref{fig:abmlForIncCovidT} shows a similar plot except
 the vertical axis is plotted on a signed square-root transformed scale to de-emphasize the
COVID period. Here is is clear that FORLAP succeeds 16 times out of 20 (80\%), whereas
Box-Jenkins succeeds 13 times (65\%).

After extensive experiments with the ABML data it seems that Haar wavelets perform best. Also,
FORLAP's advantage persists over high step-ahead forecasts (e.g.\ $h=2, 3$), but diminishes and becomes
worse
the further into the future. This is perhaps not surprising, due to the quite strong nonstationarity and
Box-Jenkins, in this case, still providing some degree of `catch all' forecast accuracy.

\section{Discussion}\label{sec:discussion}
We have demonstrated how the recently proposed local partial autocorrelation
function  introduced in \cite{killick19:lpacf} can be used to select the $p$ parameter, i.e. how much recent data is relevant, in the locally stationary forecasting method of \cite{fryz03:forecasting}.  Our subsequent modified forecasting method, $\us$,  outperforms not only the original  \cite{fryz03:forecasting} approach, but also the method based on direct time-varying autoregressive estimation and the Box-Jenkins method, on the majority of our simulated examples and on the practical data application.

A bootstrap test might provide an alternative to using asymptotic-based confidence intervals to select $\hat{p}$.
One could view $p$ as a
kind of local autoregressive model order, even if the underlying process was not autoregressive, and attempt to develop a theory of how well
$\hat{p}$ estimates it if the underlying process was autoregressive, but this is beyond the scope of the current work.
Additionally, the simulation results in Section~\ref{sec:simulations} provide good examples of
the phenomenon highlighted by~\cite{kley17:predictive}, where the best
predictor of a nonstationary time series might be a stationary predictor, and vice versa,
e.g.\ Model~E in Table~\ref{tab:epicrns}. Finally, the predictive interval coverage, and
our use of the least-squares loss function are not necessarily the only, or the best, methods
for assessing predictive performance and problem-specific/problem-tailored measures are of
value. For a more detailed discussion see \cite{gneiting11:making}, for example.

For the ABML time series, it is fascinating to see how well FORLAP does against
Box-Jenkins even during periods of significant change, as, for example, during the COVID pandemic.
We can only put this down to the flexible modelling and forecasting afforded by a method that
explicitly acknowledges the nonstationary nature of the time series. Part of FORLAP's success is
that it can position its forecast intervals better, but it can respond much quicker to variance changes
and this can be seen in the forecast intervals shown in Figures~\ref{fig:abmlForIncCovid}
and~\ref{fig:abmlForIncCovidT}.

Further research is also necessary to properly develop a mature understanding of the
proposed forecasting methodology, particularly on comparisons with other methods for forecasting of nonstationary series (where code is not freely available), on different types of series, and at forecast horizons other than $h=1$.

The $\us$ forecasting method is available within the \texttt{forecastLSW} \texttt{R} package on CRAN.

\section*{Acknowledgments}

The authors were partially supported by the Research
Councils UK Energy Programme. The Energy Programme is an RCUK
cross-council initiative led by EPSRC and contributed to by ESRC, NERC,
BBSRC and STFC. GPN gratefully acknowledges support from EPSRC
grant K020951/1.

\bibliographystyle{rss.bst}
\bibliography{bibsfull}

\appendix
\section{Simulation study models for nonstationary series}\label{app:sim.models}

Model D corresponds to the time-varying autoregressive TVAR$(1)$
model $X_t = \alpha_t X_{t-1} + Z_t$ for $t=1, \ldots, 128$
and $\alpha_t = \alpha(t/T)$ where $\alpha(z) = 1.8 z - 0.9$ for $z \in (0,1)$.

Model E corresponds to the  TVAR$(1)$ model with the same specification as for Model D except that
\begin{displaymath}
\alpha(z) = \begin{cases}
5.6z - 0.9 & \text{for $z \in (0, 1/8)$},\\
4.8z - 0.8 & \text{for $z \in (1/8, 2/8)$},\\
3.2z - 0.4 & \text{for $z \in (2/8, 3/8)$},\\
0.8 & \text{for $z \in (3/8, 5/8)$},\\
-2.4z + 2.6 & \text{for $z \in (5/8, 6/8)$},\\
-7.2z + 5.4 & \text{for $z \in (6/8, 7/8)$},\\
-1.6z + 0.5 & \text{for $z \in (7/8, 1)$}.
\end{cases}
\end{displaymath}

Model F corresponds to the TVAR$(2)$ model $X_t = \alpha_{1, t} X_{t-1} + \alpha_{2, t} X_{t-2} + Z_t$
with $\alpha_{i,t} = \alpha_i (t/T)$ for $i=1, 2$ and $\alpha_1 (z) =  \alpha_2(z) = 1.6z - 1.1$ for $z\in (0,1)$.

Model G corresponds to the TVAR$(12)$ model
$X_t = \alpha_{1, t} X_{t-1} + \alpha_{2, t} X_{t-2} + \alpha_{12, t} X_{t-12} + Z_t$ where
$\alpha_{i, t} = \alpha_i (t/T)$ for $i = 1, 2, 12$ and $\alpha_1(z) = \alpha_2(z) = 0.7z - 0.4$
and $\alpha_{12}(z) = 0.3z$ for $z \in (0,1)$.

Model H corresponds to the TVMA$(1)$ model $X_t = Z_{t} + \beta_t Z_{t-1}$ where
$\beta_t = \beta(t/T)$ where $\beta(z) = 1$ for $z\in (0, 0.9)$ and $\beta(z) = -1$ for $z \in (0.9, 1)$.

Model I corresponds to the TVMA$(1)$ model as in Model~H but with
$\beta(z) = 2z - 1$ for $z \in (0,1)$.

Model J corresponds to the TVMA$(2)$ model $X_t = Z_t + \beta_{1,t} Z_{t-1} + \beta_{2, t} Z_{t-2}$
with $\beta_{i, t} = \beta_i (t/T)$ for $i= 1,2$ where $\beta_1(z) = 2z - 1$ and
$\beta_2(z) = 9z - 0.8$ for $z\in (0,1)$.

Model K corresponds to uniformly modulated white noise, see \citet[page 826]{priestley82:spectral},  where
$X_t = \sigma^2_t Z_t$ where $\sigma_t^2 = \sigma^2 (t/T)$ where $\sigma^2(z) = (9z+1)^{3/2}$ for
$z\in (0,1)$. Note, this model might be a good fit for the ABML time series from  Section~\ref{sec:data}.

Model L is a locally stationary wavelet (LSW) process  from \cite{nason00:wavelet} with spectrum P3
from \cite{nason13} defined by $S_j(z) = 0$ for $j>2$, $S_1(z) = \frac{1}{4} - (z - \frac{1}{2})^2$
and $S_2(z) = S_1 ( z+ \frac{1}{2})$ for $z\in (0,1)$, assuming periodic boundaries for spectrum construction only.

Model M is a LSW process from \cite{nason00:wavelet} with spectrum P4
from \cite{nason13} defined by $S_j (z) = 0$ for $j=2, j>4$ and
$S_1(z) = \exp \left\{ -4(z - \frac{1}{4})^2 \right\}$, $S_3(z) = S_1(z - \frac{1}{4})$ and
$S_4(z) = S_1(z + \frac{1}{4})$, for $z\in (0,1)$, assuming periodic boundaries for spectrum construction only.
The process, $X_t$, is computed for $t=1, \ldots, 512$ and the first $T=350$ values are returned.	

\end{document}